\documentclass[conference,dvipsnames]{IEEEtran}
\usepackage[utf8]{inputenc}
\usepackage{setspace}
\usepackage{epsfig}    
\usepackage{amsfonts}
\usepackage{amssymb}
\usepackage{amsmath}
\usepackage{comment}
\usepackage{multirow}
\usepackage{psfrag}
\usepackage[linesnumbered,ruled]{algorithm2e}
\usepackage{soul} 
\usepackage{url}

\usepackage{float}
\usepackage{tikz}
\usetikzlibrary{arrows.meta, decorations.markings}
\usetikzlibrary{automata}
\usepackage{lipsum}  
\usepackage{subcaption}
\usepackage{xcolor}

\newcommand{\EG}[0]{$\varepsilon$-greedy }
\newcommand{\ES}[0]{$\varepsilon$-sticky }
\newcommand{\E}[0]{$\varepsilon$ }
\begin{document}

\title{Decentralized AP selection using Multi-Armed Bandits: Opportunistic $\varepsilon$-Greedy with Stickiness}

\author{Marc Carrascosa, Boris Bellalta \\
Wireless Networking Research Group, Universitat Pompeu Fabra\\
Email: \{marc.carrascosa, boris.bellalta\}@upf.edu
}

\date{}

\maketitle

\begin{abstract}

WiFi densification leads to the existence of multiple overlapping coverage areas, which allows user stations (STAs) to choose between different Access Points (APs). The standard WiFi association method makes the STAs select the AP with the strongest signal, which in many cases leads to underutilization of some APs while overcrowding others. To mitigate this situation, \textit{Reinforcement Learning} techniques such as \textit{Multi-Armed Bandits} can be used to dynamically learn the optimal mapping between APs and STAs, and so redistribute the STAs among the available APs accordingly. This is an especially challenging problem since the network response observed by a given STA depends on the behavior of the others, and so it is very difficult to predict without a global view of the network.

In this paper, we focus on solving this problem in a decentralized way, where STAs independently explore the different APs inside their coverage range, and select the one that better satisfy its needs. To do it, we propose a novel approach called Opportunistic $\varepsilon$-greedy with Stickiness that halts the exploration when a suitable AP is found, then, it remains associated to it while the STA is satisfied, only resuming the exploration after several unsatisfactory association periods. With this approach, we reduce significantly the network response variability, improving the ability of the STAs to find a solution faster, as well as achieving a more efficient use of the network resources. 
  
\end{abstract}



\section{Introduction}


WiFi networks are ubiquitous nowadays, and the demand for higher data rates and area coverage keeps increasing, as well as the amount of wireless devices per user. Wired traffic accounted for 50\% of the Internet traffic in 2015, but it is expected to account only for the 33\% of it by 2020, with WiFi increasing from 42\% to 49\%. This increase in the popularity of WiFi can also be seen in the number of public hotspots around the world. There were 94 million hotspots in 2016, and it is expected to reach 542 million by 2021 \cite{cisco1}. 

Network densification by deploying more APs as a way of coping with the increasing traffic demands is leading to multiple overlaps between AP's coverage areas. This densification is extending to all types of deployments, from households to public spaces in cities, where in all cases multiple APs are deployed to cover entirely the area. To deal with this densification, the new IEEE 802.11ax amendment will offer solutions addressing specifically these kind of scenarios \cite{7422404}. 


The standard association for IEEE 802.11 networks uses the Strongest Signal (SS) method to associate a user station (STA) to an AP. It scans the spectrum for all possible available networks and chooses the one with the highest Received Signal Strength Indicator (RSSI) from the received beacons. This method can lead to uneven loads by overcrowding a single AP and leaving others underused \cite{Balachandran:2002:CUB:511334.511359}, thus dense WLANs with multiple APs are in need of new association schemes that leverage such a situation, distributing the STAs among the available APs in a way that maximizes the quality of the users experience. 

AP selection and load balancing have been extensively studied as a way to improve network throughput. In \cite{1313270} a scheme is proposed in which neighboring APs compare their traffic loads to decide if they should force the disassociation of a user so that it reassociates to an underloaded AP. The authors in \cite{4151354} use the delay between a \textit{probe request} being sent and a \textit{probe response} being received as a measure of the load of the AP, and base their association scheme on picking the AP with the lowest delay instead of the lowest RSSI. In \cite{4796200} the authors use cell breathing techniques to balance the load among APs by modifying the transmission power of the \textit{beacons} sent by the AP, virtually reducing their coverage area so that STAs reassociate to other uncongested APs. A solution based on inter-AP interference is proposed in \cite{7876094}, where the STAs estimate the Signal to Interference plus Noise Ratio (SINR) from interfering APs by sending \textit{probe requests} to all APs. Then, from the \textit{probe responses} received, they can estimate the SINR, and choose the best one to find the optimal association for each STA.


In \cite{6181180}, the authors propose the use of a decentralized neural network with a single hidden layer that uses the SNR, number of STAs detected, probability of retransmissions and channel occupancy as inputs to predict the throughput achievable for each AP in the network, as well as the  optimal association to the one that maximizes it. To the best of our knowledge there are no other papers in the area of RL applied to user association. The use of MABs however is starting to be familiar to solve optimization problems in decentralized and complex scenarios. For example, the authors in \cite{7792374} give an overview of the multi armed bandits problem, as well as its applications in wireless networks as a way to solve resource allocation issues. The work in \cite{DBLP:journals/corr/abs-1710-11403} uses several Reinforcement Learning algorithms to find the optimal selection of channel and transmission power for each AP in a network. In \cite{6939716} the authors use MABs in device to device communication systems to help users choose the optimal channel and improve their performance. 

\begin{figure*}[ht]
\centering
   \begin{subfigure}[b]{.4\textwidth}
        \includegraphics[width=\textwidth]{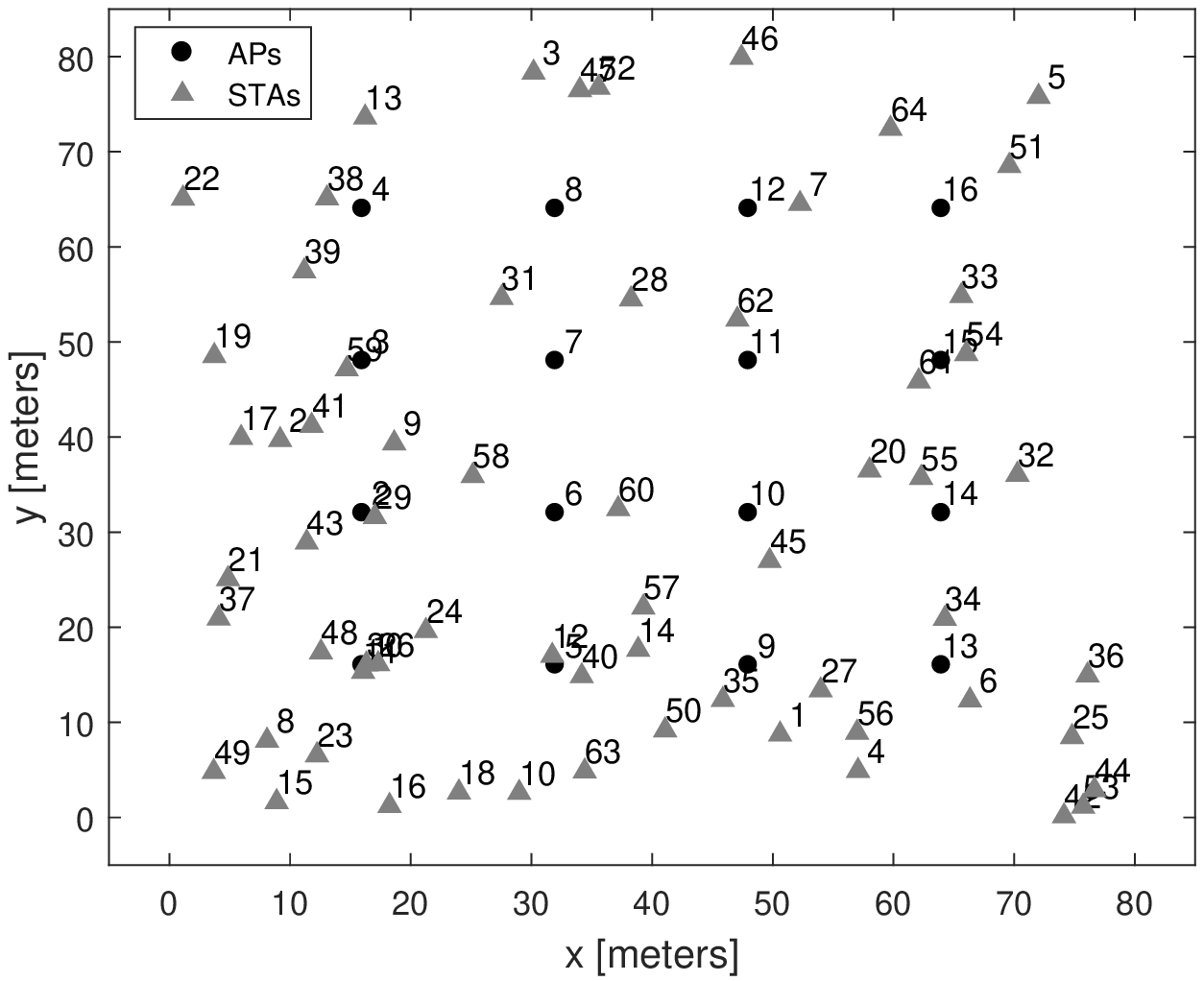}
    \caption{Scenario with the STAs distributed uniformly at random.}
    \label{scenRand}
    \end{subfigure}
    \begin{subfigure}[b]{.4\textwidth}
        \includegraphics[width=\textwidth]{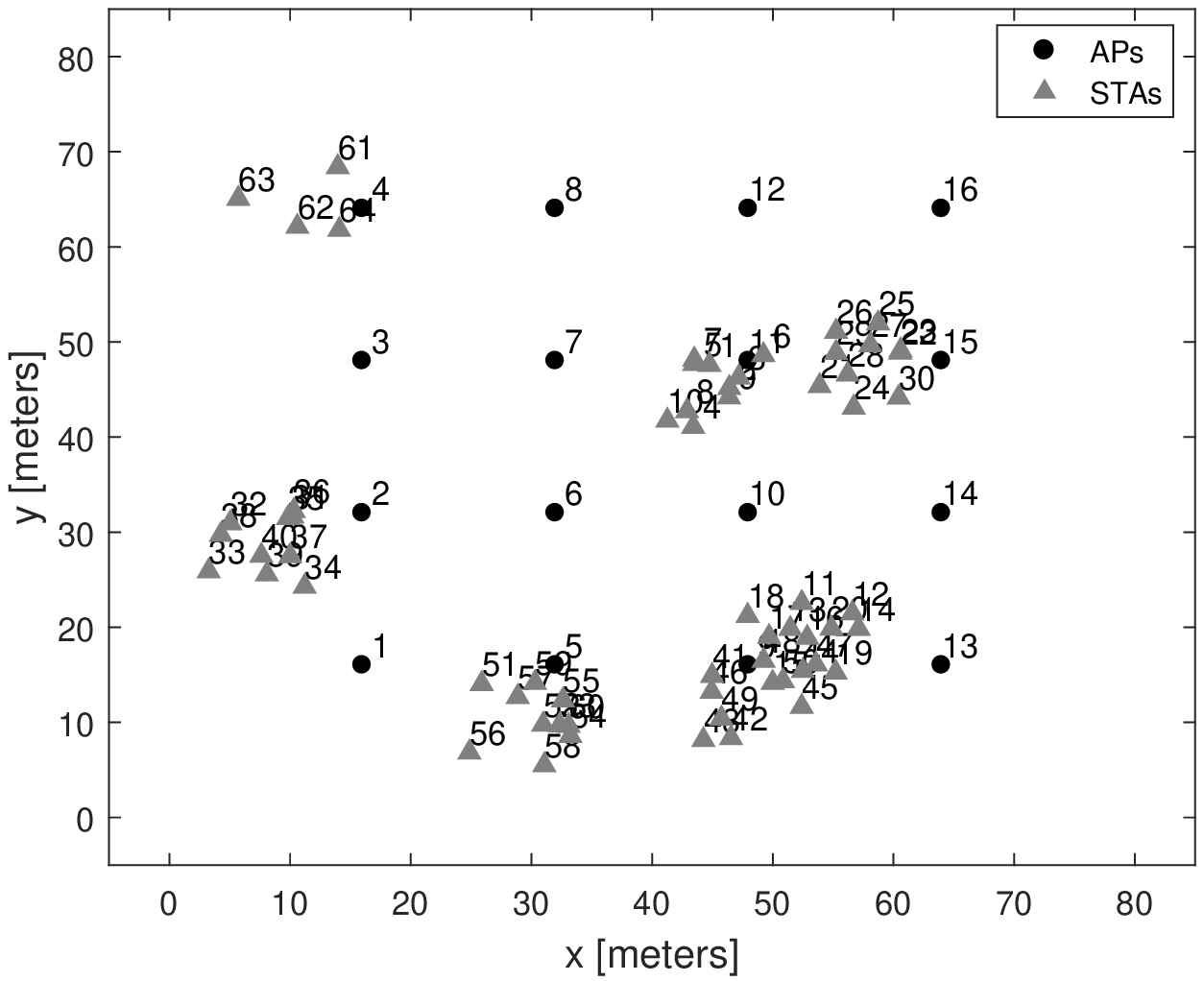}
    \caption{Scenario with the STAs grouped in clusters. Clusters are generated uniformly at random.}
    \label{scenClust}
    \end{subfigure}
     \caption{ Snapshot of the scenarios considered in this paper.}
     \label{Fig:Scenarios}
\end{figure*}
In this work, our aim is to evaluate the suitability of using Reinforcement Learning to improve the network performance by finding a feasible AP-STA association. In particular, we model the AP-STA association problem as a MAB problem, in which an agent placed in each STA can take multiple actions (i.e., AP selected), and needs to find a way to maximize its rewards by exploring them, learning more about the network at each step, and exploiting the most suitable alternative. The main challenge in our scenario is that the response obtained for each action also depends on the actions taken by the other STAs, which are completely independent, and thus, choosing the same action at different time instants may result in different outcomes, significantly increasing the action's uncertainty. To this effect, we introduce the Opportunistic $\varepsilon$-greedy algorithm with Stickiness. It follows the default exploration-exploitation tradeoff of the basic $\varepsilon$-greedy algorithm, but it includes two other features: 1) It is opportunistic in the sense it halts the exploration when it finds a satisfactory AP, and 2) When an AP becomes unsatisfactory, STAs stick to it for SC consecutive unsatisfactory association periods before exploring other APs. This approach aims to enhance the convergence speed by reducing the number of STAs changing at the same time, and remove unnecessary reassociations due to the behavior of the others.


The rest of the paper is structured as follows. Section \ref{sysMod} explains the system model used. Section \ref{SelMod} introduces the algorithms used for the AP selection. Section \ref{PerEval} presents the results obtained in the different experiments. A final summary can be found in Section \ref{Conc}, as well with several future research directions.


\section{System Model}\label{sysMod}

We consider a network deployed in a given area that consists of $M$ APs and $N$ stations (Figure \ref{Fig:Scenarios}). STAs are always active and require a throughput $w$ bps to be satisfied. We assume all the traffic in the network is downlink. STAs are equipped with an agent in charge of selecting the AP to use taking into account if the station's required throughput is achieved. Decisions about remaining in the same AP or selecting a new one are done by the $N$ STAs at every reassociation period. We assume the time between two consecutive reassociation periods is large enough (i.e., every 1 or 2 minutes) to make negligible the IEEE 802.11 association overheads, and have enough time to assess the received service\footnote{Reassociation from one AP to another can take up to $500$ ms in IEEE 802.11b devices \cite{Mishra:2003:EAI:956981.956990}, and less than $100$ ms in IEEE 802.11r compliant ones~\cite{4563250}.}. All APs operate in the 5 GHz band, using 20 MHz channels. A total of $8$ available channels is considered. Each AP chooses one of the available channels uniformly at random. The AP transmission power is set to $P_t= 20$ dBm.

In the following subsections we introduce the path-loss model considered, and detail how the required air-time per station is calculated, as well as the station's satisfaction metric used to evaluate the different options. The notation used is summarized in Table \ref{teVar}, including their values later used in Section \ref{PerEval}.


\subsection{Path-loss and transmission rate selection}

The path-loss between the APs and the STAs is obtained using the 5GHz TMB model for indoors \cite{tmb}, It is given by:
	\begin{equation}
    \begin{split}
		\text{PL}_{\text{TMB}}(d_{i,j})= L_0 + 10\gamma\log_{10}(d_{i,j}) + k\overline{W} d_{i,j} + G_s
	\end{split}
\end{equation}
where $d_{i,j}$ is the distance between STA $i$ and AP $j$, $k$ is the wall attenuation factor, and $\overline{W}$ is the average number of traversed walls per meter. $\text{G}_s$ is a random variable uniformly distributed modelling the shadowing. For all those parameters, the same values as in \cite{tmb} are used.
\setlength\tabcolsep{3 pt}
	\begin{table}[ht]\centering
	\caption{Notation used}
	\begin{small}
    \begin{tabular}{p{4cm}|c|c}
  		\hline  
  		 
		\textbf{Name}  &   \textbf{Variable}& \textbf{Value}\\ \hline\hline
	
		Legacy preamble  & $T_{\text{PHY-legacy}}$ & $20 \mu s$\\
		HE Single-user preamble & $T_{\text{PHY-HE-SU}}$ & $52 \mu s$\\
		OFDM symbol duration& $\sigma$ & $16 \mu s$\\
		OFDM Legacy symbol dur. & $\sigma_{\text{Legacy}}$ & $4 \mu s$\\
        Short InterFrame Space & SIFS & $16 \mu s$\\
        DCF InterFrame Space & DIFS & $34\mu s$\\
		Average back-off duration  & $E[\psi]$ & 7.5 slots\\        
        Empty backoff slot & $T_e$ & $9 \mu s$\\ \hline
		Service Field   &  $L_{\text{SF}}$ & 32 bits\\
		MAC header   & $L_{\text{MH}}$ & 272 bits\\
		Tail bits & $L_{\text{TB}}$& 6 bits\\ 
		ACK bits & $L_{\text{ACK}}$ & 112 bits \\
		Frame size &  $L$ & 12000 bits\\ 
		\hline

	\end{tabular}
	
	\label{teVar}
	\end{small}
	\end{table}

Using the obtained $\text{PL}_{\text{TMB}}$ values, we obtain the transmission rate used for the communication between each AP-STA pair, i.e. $P_r = P_t -\text{PL}_{\text{TMB}}(d_{i,j})$. Then, using the received power as a reference, we obtain both the transmission rate, $r$, and the legacy transmission rate $r_L$.


\subsection{Required airtime per STA}

The required airtime per STA and per second is calculated taking into account the throughput required by a station, $w$, the average packet sizes it transmits, $L$, the transmission rate $r$, and all other IEEE 802.11 overheads. In detail, the duration of a transmission for STA $i$ is given by:

\begin{align}
     T(r_i,r_{L,i}) = T_{\text{data}}(L,r_i) + \text{SIFS} + T_{\text{ack}} +  \text{DIFS} + T_e
\end{align}
where
\begin{align}
    T_{\text{data}}(r_i) = T_{\text{PHY-HE-SU}} +  \bigg\lceil\frac{L_{\text{SF}} + L_{\text{MH}} + L_i + L_{\text{TB}}}{r_i} \bigg\rceil \sigma
\end{align}
and
\begin{align}
     T_{\text{ack}}(r_{L,i}) =  T_{\text{PHY-legacy}}  + \bigg \lceil \frac{L_{\text{SF}} + L_{\text{ACK}}+ L_{\text{TB}}}{r_{L,i}} \bigg \rceil    \sigma_{\text{legacy}}
\end{align} 
Then, the airtime required by STA $i$ is given by

\begin{align} \label{eq:flow_util}
	u(\omega,L,r_i,r_{L,i})=\frac{w}{L} \cdot (E[\psi] T_e + T(L,r_i,r_{L,i}))
\end{align}


\subsection{Airtime Occupancy per AP}

The airtime channel occupancy observed by AP $j$ is given by 

\begin{align} \label{eq:ap_util}
	U_{j}=\min(1,\sum_{\forall i \in \mathcal{S}_j}{u(\omega,L,r_i,r_{L,i})})
\end{align}
where $\mathcal{S}_j$ is the set of all stations associated to AP $j$ and to other APs within the coverage range of AP $j$ that operate in the same channel.



\section{MAB-based AP-selection mechanism}\label{SelMod}


In this section, we describe the AP-selection mechanism presented in this paper. The initial association is done with the SS method, in which the STAs choose the AP with the strongest signal out of the ones they can perceive. Afterwards, they use the $\varepsilon$-greedy or the $\varepsilon$-sticky algorithm to reassociate to other APs in range with the aim to improve their satisfaction.

\subsection{AP-selection using baseline $\varepsilon$-greedy}

\begin{figure}[ht]
\centering
        \includegraphics[width=.49\textwidth]{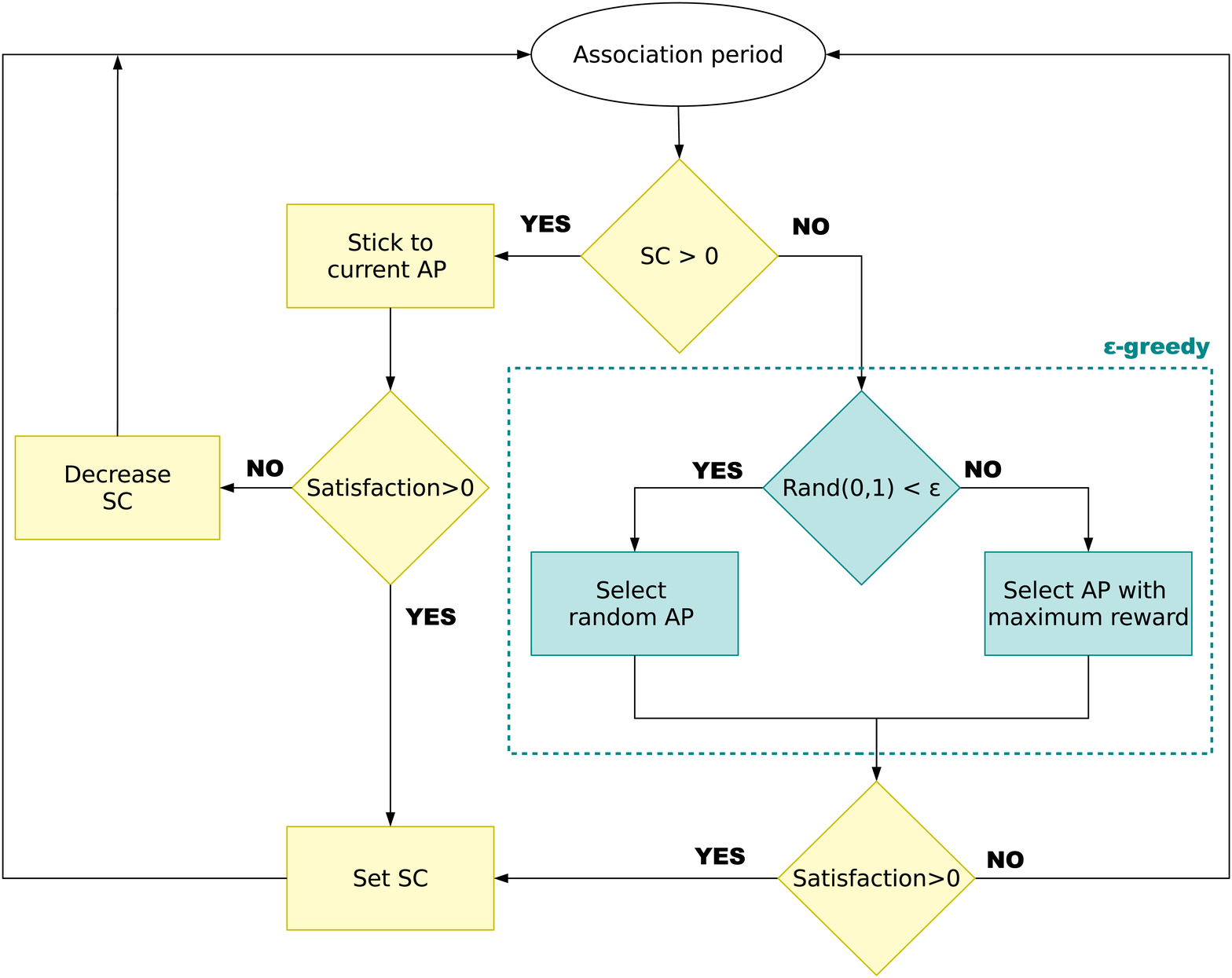}
    \caption{Block diagram for \ES}
    \label{stickyblocks}
\end{figure}

Each STA keeps a list with all the APs it is able to detect and their accumulated reward. The value of $\varepsilon$ dictates how often the STA will explore the system or exploit the information that it has already acquired. If the STA explores, it re-associates to a random AP from the ones in its list. If it exploits, the STA picks the AP with the highest accumulated reward. Every action taken receives a reward between $0$ and $1$. Figure \ref{stickyblocks} shows the decision flowchart for \EG in green.

\begin{figure*}[ht]
\centering
   \begin{subfigure}[b]{.45\textwidth}
        \includegraphics[width=\textwidth]{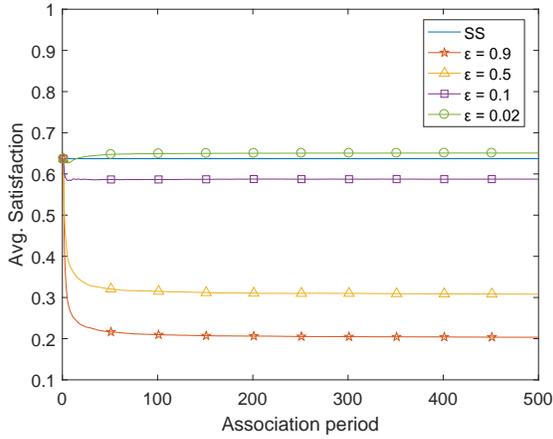}
    \caption{ $\varepsilon$ values for \EG  with random STA placements}
    \label{eg}
    \end{subfigure}
    \begin{subfigure}[b]{.45\textwidth}
        \includegraphics[width=\textwidth]{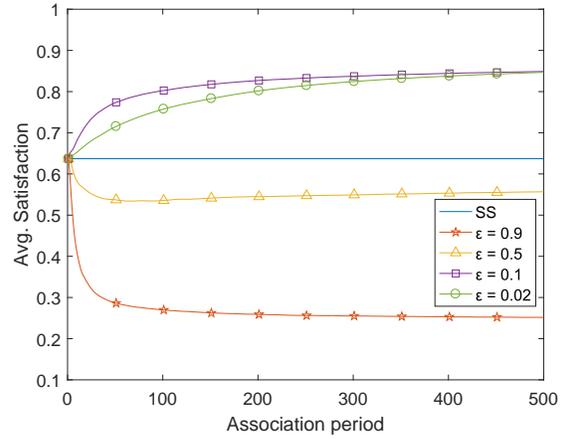}
    \caption{ $\varepsilon$ values for $\varepsilon$-sticky with random STA placements}
    \label{es}
    \end{subfigure}
   \begin{subfigure}[b]{.45\textwidth}
        \includegraphics[width=\textwidth]{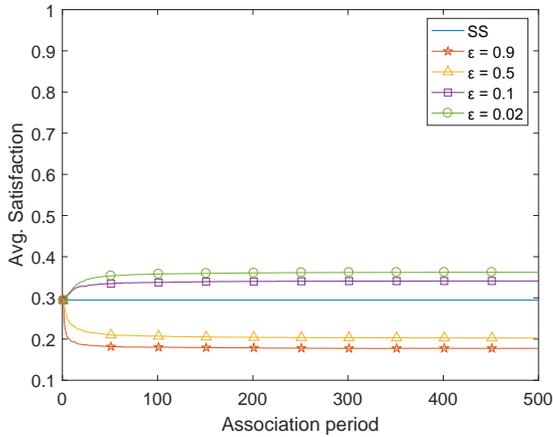}
    \caption{$\varepsilon$ values for $\varepsilon$-greedy with clustered STA placements}
    \label{egc}
    \end{subfigure}
   \begin{subfigure}[b]{.45\textwidth}
        \includegraphics[width=\textwidth]{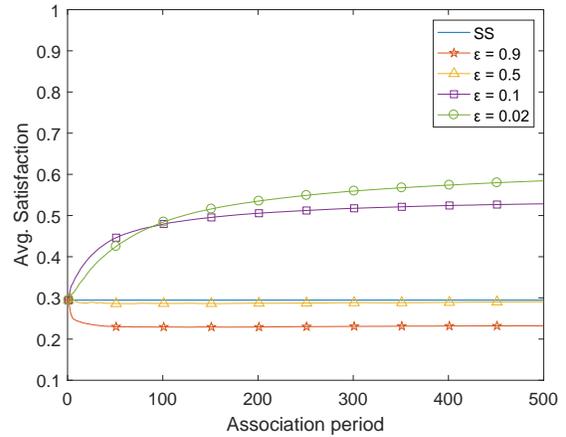}
    \caption{$\varepsilon$ values for $\varepsilon$-sticky with clustered STA placements}
    \label{esc}
    \end{subfigure}
     \caption{Selection of optimal $\varepsilon$ values}
     \label{epsilonSelection}
\end{figure*}

\subsection{AP-selection using Opportunistic $\varepsilon$-greedy with stickiness}

In order to improve the performance of the \EG algorithm, we extend it by including stickiness, i.e., once a STA has found an AP that satisfies its requirements, it will remain associated to it, and will only restart exploring other APs after SC consecutive unsatisfactory association periods. Following this approach, the STA avoids exploring needlessly if its satisfaction is only temporarily affected by the exploration of other STAs, as well as in those cases where it has already found a suitable solution. Figure \ref{stickyblocks} details the \ES algorithm.


\subsection{Definition of the reward}

The reward that STA $i$ gives to AP $j$ is the airtime received by the STA, which is given by  \eqref{Eq:satisfaction2}. Therefore, if the network can accommodate the required airtime, then the STA is considered to be satisfied, and the reward for the current AP is increased by one.

\begin{align} \label{Eq:satisfaction2}
	\zeta_{i}=\frac{u_{i}(\omega,L,r_z,r_{L,z})}{\max(1,\sum_{\forall z \in \mathcal{S}_j}{u(\omega,L,r_z,r_{L,z})})}
\end{align}


\section{Performance Evaluation}\label{PerEval}

In this section we compare the impact of the previous two \EG algorithms on the STA's satisfaction, showing its temporal evolution when the stations are uniformly distributed or grouped in clusters. We study the optimal \E and sticky counter values, as well as the effects of increasing the number of APs and the required throughput of the STAs. The parameters used and their values are shown in Table \ref{teVar}. Each simulation is repeated 100 times, and the results presented are the average of all simulations. All the code used can be found in our github repository. \footnote{\scriptsize{\url{https://github.com/wn-upf/Decentralized-AP-selection-using-Multi-Armed-Bandits}}}


\subsection{Toy scenario with different STA distributions}

We start by studying the effect of the STA distribution on the performance of the algorithms. To do this we will use a toy scenario with fixed AP positions to better compare the random and clustered STA distribution, and better define the different algorithm configurations.

We set $16$  APs in a $4x4$ grid in a square area of $80x80$ metres. Each AP selects one out of eight channels at random, and we use two different distributions for the placement of $64$ STAs. In the first one we place them randomly following a uniform distribution, and in the second one we set clusters of $10$ STAs distributed along areas of $10x10$ metres with the center of each cluster being chosen at random. Each STA requests $4$ Mbps.

\begin{figure*}[ht]
\centering
   \begin{subfigure}[b]{.45\textwidth}
        \includegraphics[width=\textwidth]{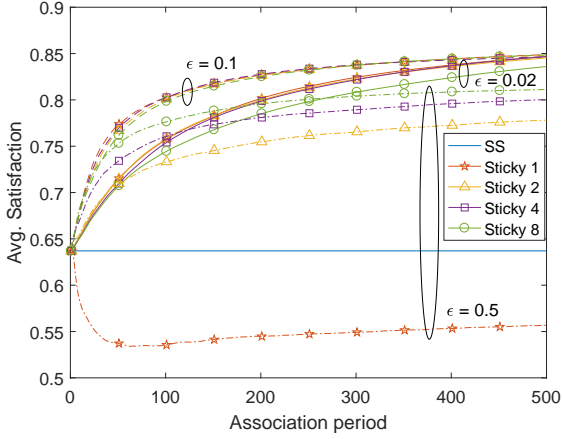}
    \caption{ Sticky counter values for random STA placements}
    \label{egcl}
    \end{subfigure}
    \begin{subfigure}[b]{.45\textwidth}
        \includegraphics[width=\textwidth]{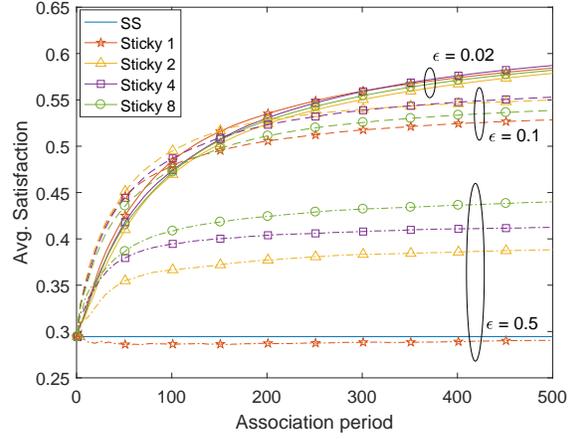}
    \caption{Sticky counter values for clustered STA placements}
    \label{escl}
    \end{subfigure}
     \caption{Selection of optimal sticky values }
     \label{epsilonSelectioncl}
\end{figure*}

Figure \ref{epsilonSelection} shows the average satisfaction per iteration (i.e., the accumulated satisfaction normalized by the number of iterations elapsed, and averaged over all STAs) obtained by each algorithm for each value of $\varepsilon$. For the case where STAs are uniformly distributed, using \EG we can observe in Figure \ref{eg} that we can only improve upon SS with $\varepsilon = 0.02$. Higher $\varepsilon$ values cannot compete with the SS method, as it seems that they do not learn properly. Further, even when using $\varepsilon = 0.02$ we only obtain a $2.1\%$ improvement. This changes for the clustered environment in Figure \ref{egc} however, where we obtain a $15.8\%$ and $23.02\%$ improvement when using $\varepsilon = 0.1$ and $\varepsilon = 0.02$ respectively. This is due to the fact that when the STAs are uniformly distributed, they are spread evenly among the APs, while for the clustered distribution multiple STAs are placed close to the same AP, thus selecting it, and leaving others underused. As it can be observed, \EG algorithm is then capable of balancing the network's load by distributing the STAs between all APs.

Next, we try our \ES algorithm for both cases, using a sticky counter of 1. Figure \ref{es} shows the results for the uniform distribution, where we can observe a higher improvement over the \EG results (Figure \ref{eg}). Now, we obtain a $33.26\%$ and $32.9\%$ improvement over SS with $\varepsilon = 0.1$ and $\varepsilon = 0.02$, respectively. For the clustered STA distribution, Figure \ref{esc} shows even better results, with a gain of $79.45\% $ for $\varepsilon = 0.1$ and of $98.43\%$ for $\varepsilon = 0.02$.

Based on the presented results, we can conclude the following: first, we need a low exploration rate to obtain good results. This is due to the fact that high exploration rates lead to a high variability, meaning that the information obtained by the STAs in past iterations is irrelevant for the next association period. With a low exploration rate, only a few STAs select a different AP at each association period, so the scenario remains fairly stable,  allowing the STAs to keep up with the changes in the network.

This can also be observed when \ES is used, where the stickiness keeps the exploration rate even lower, allowing the information learned by the STAs to be more relevant. Another aspect to mention, especially for the \ES case, is that using $\varepsilon = 0.1$ in the first association periods leads the STAs to learn at a faster rate than using $\varepsilon = 0.02$,  but using $\varepsilon = 0.02$ leads to a higher slope on the learning curve, meaning that $\varepsilon = 0.02$ will lead to higher satisfaction with enough association periods. This is especially visible in Figure \ref{esc}, where $\varepsilon = 0.1$ and $\varepsilon = 0.02$ intersect around iteration 100 and, after that, $\varepsilon = 0.02$ continues increasing at a faster rate. Finally, we can observe that both \EG and \ES work better when STAs are grouped in clusters, where SS performs badly.

\begin{figure}[t]
    \centering
    \includegraphics[width=0.45\textwidth]{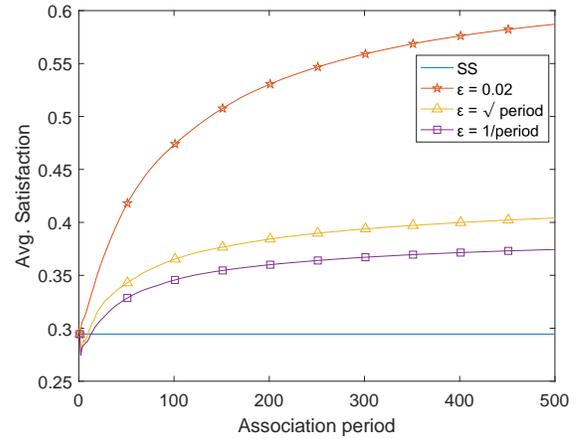}
    \caption{$\varepsilon$ comparison }
    \label{rewcalc}
\end{figure}

\begin{figure*}[ht]
\centering
   \begin{subfigure}[b]{.45\textwidth}
        \includegraphics[width=\textwidth]{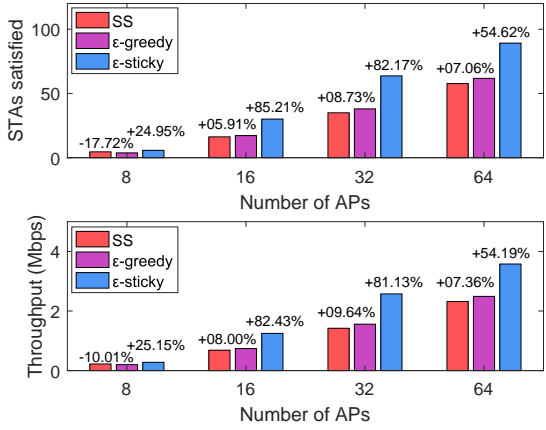}
    \caption{STAs satisfied and Throughput achieved when increasing the number of APs}
    \label{apinc}
    \end{subfigure}
    \begin{subfigure}[b]{.45\textwidth}
        \includegraphics[width=\textwidth]{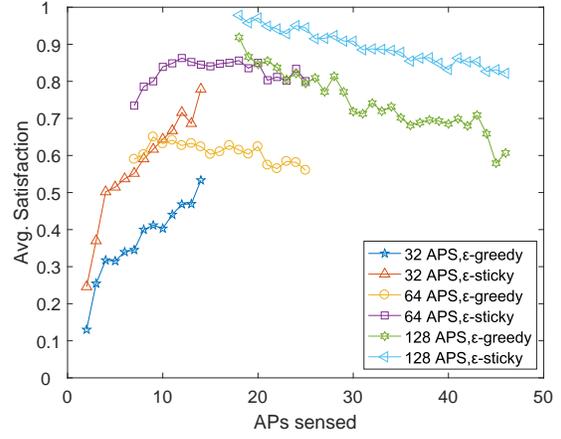}
    \caption{Satisfaction vs. average APs sensed by STAs (100 samples minimum)}
    \label{lastm}
    \end{subfigure}
     \caption{Effects of increasing the number of APs }
     \label{incs}
\end{figure*}	

We have been using \ES with a sticky counter of $1$, meaning that as soon as a STA is dissatisfied once it returns to the \EG behaviour. We now show the effects of changing the value of the sticky counter and analyze how the satisfaction obtained by the STAs changes. Figure \ref{epsilonSelectioncl} shows the average satisfaction per iteration for $\varepsilon$ values of $0.02, 0.1$ and $0.5$ with increasing values for the sticky counter. Each $\varepsilon$ is plotted with a solid, dashed or dash-dotted line respectively. In the case where STAs are uniformly distributed (Figure \ref{egcl}), we can observe that for low $\varepsilon$ values the sticky counter does not have a significant effect, but for $\varepsilon = 0.5$ it goes from a satisfaction of $55.68\%$ with a SC $ = 1$ to a satisfaction of $81.12\%$ with a SC $ = 8$ , which means going from a $31.36\%$ decrease over SS to an increase of $27.34\%$. For $\varepsilon = 0.02$ the higher performance is achieved with a sticky counter of 4 and a satisfaction of $84.68\%$ (with a sticky counter of 1 giving us $84.67\%$), and for $\varepsilon = 0.1$ a sticky counter of 1 is the best with a satisfaction of $84.89\%$. For the case in which STAs are grouped in clusters (Figure \ref{escl}), the impact of the sticky counter is more significant, showing a clear improvement with larger sticky counter values. For instance, for both $\varepsilon = 0.1$ and $\varepsilon = 0.02$, the best performance is achieved using SC $= 4$.

\subsection{Static vs decreasing $\varepsilon$}

 Some versions of the \EG algorithm use a decaying $\varepsilon$ so that the agent starts exploring to obtain a reward from all possible sources, and then exploits more and more over time. In this case however, an association from one STA to one AP has an effect on other STAs associating to that AP, meaning that if all STAs are exploring randomly no useful information is gained. A small $\varepsilon$ value limits the movement of most STAs, allowing the ones that do explore to get a good view of the current state of the network. Figure \ref{rewcalc} shows the satisfaction achieved when using the decaying \E or a static value. We use the toy scenario with \ES and a sticky counter of 4. We use two decaying methods, the first one is $\varepsilon = \frac{1}{\sqrt{\text{association period}}}$, and the second one is $\varepsilon = \frac{1}{\text{association period}}$, which decays faster than the previous one. Both methods lead to results that improve upon SS, but using a low static \E outperforms them both. 


\subsection{Increasing the number of APs}

To study the effect of increasing the number of APs in the system perfomance, we consider a scenario with $10$ clusters, distributed uniformly across a $80x80$ m area. All clusters contain $10$ STAs, and each STA requires a throughput of $w=4$ Mbps. 

Figure \ref{apinc} shows both the STAs satisfied for \EG and \ES methods as well as the throughput achieved by the end of the simulation for 8, 16, 32 and 64 APs. For 8 APs, the only case in which \EG  cannot improve the network performance, we obtain a $17.72\%$ decrease in the number of STAs satisfied, going from 4.57 with SS to 3.76 satisfied STAs with \EG,  and a $10.01\%$  decrease in the throughput achieved going from $0.22$ Mbps to $0.2$ Mbps. For every other case \EG improves upon SS, with a maximum increase obtained using $32$ APs, where we get $8.73\%$ more STAs satisfied and $9.64\%$ more throughput. Using \ES we obtain an increase for all cases, with the maximum being also on $32$ APs, with $82.17\%$ more STAs satisfied (from $34.95$ with SS to $63.67$ with \ES )and $81.13\%$ more throughput (from $1.421$ Mbps to $2.574$ Mbps). The increase of satisfaction and throughput with $64$ APs is lower than the one with $32$ APs due to the higher density of APs, leading to better results on SS, since each AP has to deal with very few STAs.

Considering that both the \EG and \ES  algorithms are based on exploring the APs in range of each STA, we can infer that the higher the number of APs sensed by a STA, the higher the potential satisfaction achievable, as more APs improve our chances to find a suitable association. Figure \ref{lastm} shows the number of association periods a STA remains satisfied according to the number of APs sensed divided by the number of considered association periods. We have considered $100$ STAs like in the previous simulation, but we have considered $32$, $64$ and $128$ APs. For $32$ APs we observe that the more APs sensed the higher the satisfaction is. For $64$ APs most STAs sense at least $7$ APs and the satisfaction only increases until they reach $12$ APs, remaining stable afterwards. For $128$ APs our STAs sense $18$ APs at the very least, and the satisfaction decreases the more APs they sense. This is probably due to the high concentration of APs (i.e., they share channels and airtime load, which leads to congestion). We can still observe however, that the best performance is achieved when a STA senses $10$ APs.

  \begin{figure}[ht]
    \centering
    \includegraphics[width=0.45\textwidth]{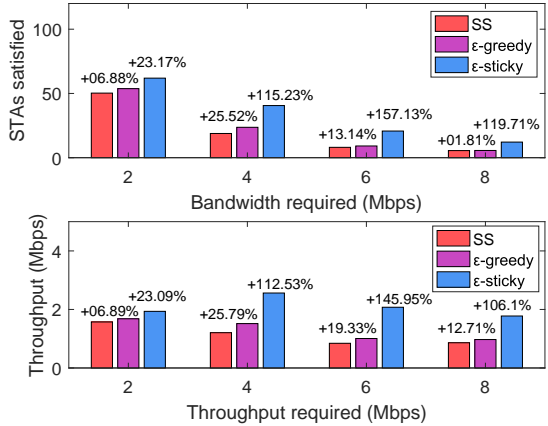}
    \caption{Effect of increasing the required throughput}
    \label{bwinc}
\end{figure}
  
\subsection{Variable throughput requirements}

Here, we investigate the effect that the throughput requirements of the STAs have on the system performance. Figure \ref{bwinc} shows the obtained results, showing both number of STAs satisfied and throughput achieved, when the throughput required by the STAs increases from $2$ to $8$ Mbps. 

When the STAs require a throughput of $2$ Mbps they become satisfied fast, as most of them are satisfied already with SS in the first association period. Then, as expected, \EG and \ES only show a small increase of $6.88\%$ and $23.17\%$ respectively. For higher throughput values (i.e., $4$, $6$ and $8$ Mbps), \EG and \ES are capable of significantly improving the system performance, with more than a $100\%$ increase in the number of satisfied STAs using \ES, with a maximum gain of $157.13\%$ for $6$ Mbps, going from $8.07$ satisfied STAs in SS to $20.75$. \EG is always successful in improving the system performance too, with a maximum gain of $25.52\%$ for $4$ Mbps.

In terms of throughput achieved, the same observations can be done, for a required throughput equal to 2 Mbps, \ES gives an average throughput per STA equal to $1.93$ Mbps, having almost all STAs satisfied. The maximum gain for \EG also appears when the required throughput is equal to $4$ Mbps, as it obtains a $25.79\%$ increase, going from $1.20$ Mbps with SS to $1.51$ Mbps with $\varepsilon$-greedy. For \ES we find the maximum gain when STAs demand $6$ Mbps, where a $145.95\%$ increase is achieved, going from $0.84$ Mbps with SS to $2.07$ Mbps with $\varepsilon$-sticky.




 	
\section{Conclusions}\label{Conc}
	
In this paper we have studied the use of \EG and \ES strategies to improve the AP-STA association process in IEEE 802.11 WLANs. We have simulated multiple scenarios to test these algorithms, finding that in environments with a high density of APs they are able to provide feasible AP-STA association solutions. Results confirm that the use of 'smart' reassociation algorithms may further improve the user's quality of experience and network utilization. 

The next challenge we want to study is the added effect of user mobility and different user profiles (i.e., streaming, web browsing, idling) on these algorithms, as well as centralized approaches in which a controller makes the reassociation decisions for the users.


\section*{Acknowledgements}\label{ACKs}

This work has been partially supported by a Gift from CISCO University Research Program (CG\#890107) \& Silicon Valley Community Foundation, by the Spanish Ministry of Economy and Competitiveness under the Maria de Maeztu Units of Excellence Programme (MDM-2015-0502), and by the Catalan Government under grant SGR-2017-1188.
\


\bibliographystyle{unsrt}
\bibliography{Bib}

\end{document}